\documentclass[floatfix,%
 aip,
 amsmath,amssymb,
 reprint,%
]{revtex4-1}

\usepackage{graphicx}% Include figure files
\usepackage{dcolumn}% Align table columns on decimal point
\usepackage{bm}% bold math
\usepackage{booktabs}
\usepackage[utf8]{inputenc}
\usepackage[T1]{fontenc}
\usepackage{mathptmx}
\usepackage{etoolbox}

\makeatletter
\def\@email#1#2{%
 \endgroup
 \patchcmd{\titleblock@produce}
  {\frontmatter@RRAPformat}
  {\frontmatter@RRAPformat{\produce@RRAP{*#1\href{mailto:#2}{#2}}}\frontmatter@RRAPformat}
  {}{}
}%
\makeatother

\newcommand{\ket}[1]{$|#1\rangle$}
\newcommand{\nth}{$\bar{n}_{\mathrm{th}}$}

\begin{document}

\title{Active isolation of transmon readout with a tunable inductive bridge}

\author{Chunyang Ding}
\email{chunyang.ding@stanford.edu}
\affiliation{Department of Physics and Applied Physics, Stanford University, Stanford, California 94305}

\author{Chuyao Tong}
\affiliation{Department of Physics and Applied Physics, Stanford University, Stanford, California 94305}

\author{Paul Varosy}
\affiliation{Department of Physics and Applied Physics, Stanford University, Stanford, California 94305}

\author{Nikitha Jain}
\affiliation{Department of Physics and Applied Physics, Stanford University, Stanford, California 94305}

\author{Ziyi Zhao}
\affiliation{Yale Quantum Institute, Yale University, New Haven, CT 06511, USA}

\author{Eric I. Rosenthal}
\altaffiliation{Present address: Sygaldry Technologies, Ann Arbor, Michigan, USA}
\affiliation{Department of Physics and Applied Physics, Stanford University, Stanford, California 94305}

\author{Konrad W. Lehnert}
\affiliation{Yale Quantum Institute, Yale University, New Haven, CT 06511, USA}

\author{David I. Schuster}
\affiliation{Department of Physics and Applied Physics, Stanford University, Stanford, California 94305}

\begin{abstract}
    Scalable superconducting quantum computation requires fast, high-fidelity readout. Currently, this is assisted through cryogenic amplification of the readout signal. In traditional dispersive readout schemes, the amplifiers that are used to provide high signal-to-noise ratio also generate thermal noise that can decohere the qubits, demanding additional ferrite-based isolators to suppress this noise. Here, we integrate a chip-scale microwave switch, the tunable inductive bridge, into the readout chain of a fixed-frequency transmon. Flux-pulsing the switch, such that it transmits only during the readout, did not make a discernable difference to qubit coherence, residual excited-state population, or readout fidelity metrics. By injecting broadband noise that mimics amplifier thermal photons, we show that the switch suppresses the resulting qubit dephasing by more than $30$~dB, consistent with its independently measured on-off ratio, establishing it as a potential chip-based integrated replacement for readout-chain isolators.
\end{abstract}

\maketitle

Quantum error correction requires high-fidelity mid-circuit measurement of many qubits~\cite{google2025quantum}, with readout resonators frequency-multiplexed across several GHz of range~\cite{chen2012multiplexed, spring2025fast}. Dispersive readout has become faster and more accurate through near-quantum limited parametric amplification~\cite{macklin2015near}, Purcell filters that permit larger resonator linewidths~\cite{reed2010fast, sunada2024photon}, and an improved understanding of measurement induced state transitions during readout that allows readout with larger photon numbers~\cite{sank2016measurement, kurilovich2025high, zobrist2026measurement}. The readout chain must transmit signal photons up toward the amplifiers with minimal loss, yet the same reciprocal path allows amplifier thermal noise to propagate backward, populating the resonator with  photons that dephase the qubit~\cite{bertet2005dephasing, gambetta2006qubit}. This reverse isolation is currently provided by ferrite-based isolators or circulators, which are macroscopic, magnetic components whose thermal load and stray fields obstruct scaling to millions of qubits. An alternative is active isolation by introducing a fast, chip-based cryogenic rf-switch. When transmitting, it allows photons to travel upstream from the device to our room temperature digitizer. When isolating, it guards the qubits from thermal and amplifier-added photons that travel downstream.

Past implementations of active isolation for qubit readout have taken varied approaches. One approach builds isolation into the quantum amplifier itself, providing forward gain while attenuating reverse propagating noise. This method has been demonstrated using optomechanical coupling~\cite{malz2018quantum}, SQUID-based circuits~\cite{thorbeck2017reverse}, and engineered traveling wave parametric amplifiers~\cite{lecocq2021efficient, kow2026traveling}. Another approach uses cryogenic rf-activated devices. Such switches have been demonstrated with tunable Josephson Junctions under resonant conditions~\cite{naaman2016chip, chang2020fast}, with nanowire biasing~\cite{wagner2019demonstration}, and with tunable inductances realized through chains of DC SQUIDs~\cite{pechal2016superconducting}. In all these cases, to achieve scalable multiplexed readout, such a switch should combine broad bandwidth, low insertion loss, negligible power dissipation, and switching times that are fast compared to the measurement cycle.  
 
In this work, we study a cryogenic microwave switch, the Tunable Inductive Bridge (TIB)~\cite{chapman2016general, chapman2017single, rosenthal2021efficient, zhao2023integrating, zhao2023integrating, zhao2026persistent} and its application to superconducting qubit's readout. The TIB is a Wheatstone bridge of flux-tunable inductors that reflects signals when all its arms are balanced, and transmits when the arms are unbalanced. It has been shown to have wide bandwidth performance of more than 30 dB of on/off contrast across an octave of bandwidth with the 1dB compression point at -98dBm of input power~\cite{rosenthal2021efficient}. As a chip-based device, the TIB could be directly integrated into the qubit readout package.

\begin{figure*}[htbp]
\centering
\includegraphics[width = 2\columnwidth]{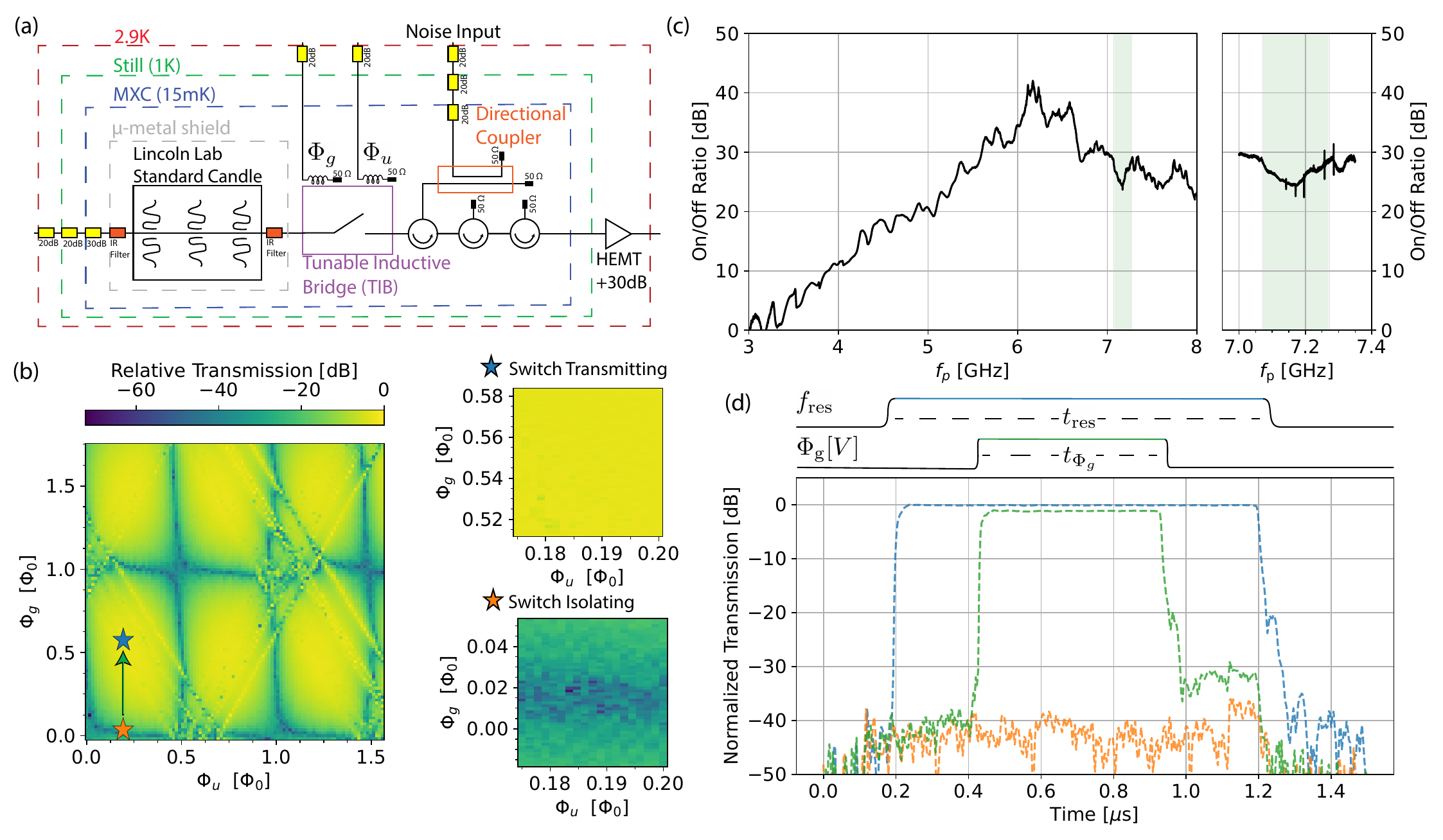}
\caption{Characterizing the rf switch as a function of its external flux biases, its frequency response, and its time-domain response. (a) The cryogenic wiring diagram for this device as measured. The switch is placed in series between the feedline output of a Lincoln Lab ``standard candle'' transmon package and a triple-junction circulator. Noise is injected via the directional coupler on the circulator, which connects to the output port of the switch. (b) Sweeping $\Phi_\mathrm{g}$ and $\Phi_\mathrm{u}$ and measuring the relative transmission through the switch at probe frequency $f_\mathrm{p} = 7.142$~GHz, referenced to the maximum transmission at this frequency. The blue (orange) star indicates the transmitting (isolating) bias location and the insets show a fine sweep near these locations. The green arrow shows the direction of the flux-pulse for activating the switch during readout. Diagonal lines may indicate flux instability. (c) The on-off transmission ratio of the switch, plotted as a function of frequency. The green shaded region is the 200MHz window near $f_\mathrm{res}$, where the on-off ratio is between 25-30dB. The inset plot is a higher resolution scan with smaller frequency steps, taken in the neighborhood of $f_\mathrm{res}$. (d) Time response data ($|S_{21}|$) of the measurement chain with an in-line TIB, using an excitation microwave pulse at $f_\mathrm{p} = 7.160$~GHz, collected with a 700MHz QICK ADC (time resolution of 2.6~ns). The pulse is sent for $t_\mathrm{res} = 1 \mu$s, as shown in the above trace. The blue and orange traces show the pulse with the switch held constantly at its transmitting and isolating bias points, respectively. For the green trace, the switch is initially held at its isolating bias, after which $\Phi_\mathrm{g}$ is pulsed to the transmitting bias for $t_\mathrm{\Phi_g} = 500$ns, following the timing diagram shown above.}
\label{fig1}
\end{figure*}

We operate the TIB as a packaged device, in a fixed-frequency transmon qubit's readout chain (Fig.~\ref{fig1}(a)), and verify its performance and its ability to suppress injected noise. The switch is placed directly before the first circulator in the outgoing readout line. We first characterize the switch's transmission properties in frequency- and time-domain measurements. Then, we implement a protocol to pulse the flux bias point of the TIB, such that it transmits only during qubit readout and is otherwise isolating, protecting the qubit from noise during its free evolution. We verify that operating the switch in this pulsed manner does not generate measurable backaction thermal photons upon the qubit, for both when the signal through the chain is at the qubit frequency and the resonator frequency. Finally, to simulate noise photons interacting with the qubit/resonator system, we bypass the circulators and inject noise at the output port of the TIB to verify its previously characterized ability in suppressing these injected photons. We measure the number of residual resonator photons via excess qubit dephasing, confirming that the switch has greater than 30dB of protection when the switch is operated in the isolating, pulsed configuration. 

\begin{figure*}[htbp]
\centering
\includegraphics[width=2 \columnwidth]{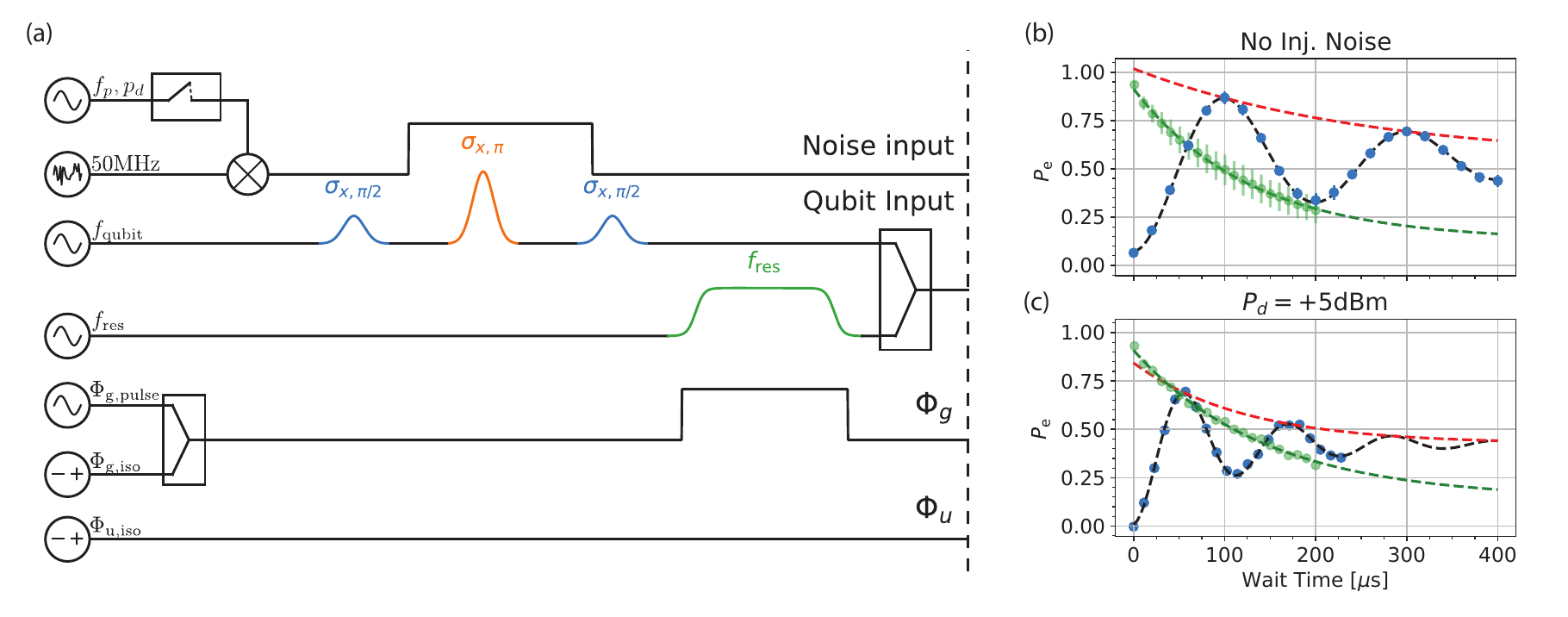}
\caption{The injected noise experiment. 
(a) The room temperature experimental setup for qubit control, switch control, and noise generation. Pulse sequences for performing the injected noise coherence experiment with a pulsed switch is shown. Note that the noise channel is gated with a room-temperature RF switch, and that the $\Phi_\mathrm{g}$ is only pulsed when the switch begins at its isolating bias. (b, c) With the switch operated in pulsed mode, echo oscillations used to extract $T_{2E}$ are shown in blue, while the excited state population decay used to extract $T_1$ is shown in green. 
The oscillations in the plot are due to phase advancement set to be two periods for the final $\pi / 2$ pulse. The black (green) dashed line indicates the fit for the decaying sinusoid (exponential decay) function; the red dashed line is the fit of the exponential envelope of the $T_{2E}$ experiment only. In (b), there is no injected noise, whereas in (c), there is injected noise with $P_\mathrm{d} = +5$~dBm. Note that we change the total length of the $T_{2E}$ experiment with $P_\mathrm{d}$, to compensate for the decreasing coherence time at higher residual photon number. The +5dBm injected noise reduces the $T_{2E}$ coherence, showing that qubit dephasing increased.}
\label{fig2}
\end{figure*}

We characterize the TIB directly in the readout chain, connected in series between the output of the qubit package and the first circulator (Fig.~\ref{fig1}a), so that the measured performance includes all losses and reflections of the assembled setup. The TIB is a Wheatstone bridge whose four arms are arrays of tunable DC-SQUIDs~\cite{chapman2016general}. For each of the identical DC-SQUIDs, the uniform flux $\Phi_\mathrm{u}$ sets the bias at its most flux-sensitive point. In contrast, the gradiometric flux $\Phi_\mathrm{g}$ lowers the inductance of each loop in one pair of arms while raising that of the other, tuning the entire switch. At $\Phi_\mathrm{g} = n \Phi_0$, the bridge is balanced and the switch reflects (isolating); when the inductance is maximally unbalanced, it transmits (transmitting). To identify the operating biases, we sweep $\Phi_\mathrm{u}$ and $\Phi_\mathrm{g}$ while measuring transmission at a probe frequency $f_\mathrm{p} = 7.1417 $ GHz, normalized to the maximum over the sweep (Fig.~\ref{fig1}b). The blue (orange) star indicates the transmitting (isolating) bias, where the relative transmission is $0 (-35)$~dB. Notably, we choose $\Phi_{u, \mathrm{iso}} = \Phi_{u, \mathrm{trans}}$, so that a single baseband voltage pulse $\Phi_\mathrm{g, pulse}$ toggles the switch (arrow in Fig.~\ref{fig1}b). We then measure transmission at both bias points, from 3 to 8~GHz, and compute the on-off ratio at each probe frequency, using a 0dBm excitation pulse and a 1~kHz RBW on a Keysight E5071 Network Analyzer (Fig.~\ref{fig1}c). Over 5--8~GHz, the typical band for readout resonators, the on-off ratio is between 20--40~dB, reaching a maximum of 42~dB at 6.17~GHz.  Within a 200~MHz window around the readout resonator studied here at $f_\mathrm{res}= 7.1717$~GHz, it is 25--30~dB.

Finally, we calibrate the time-domain response of the switch, in response to a baseband voltage pulse sent via $\Phi_{\mathrm{g}}$. Fig.~\ref{fig1}d shows the transmitted amplitude of a 1.5~$\mu$s time trace of a 1~$\mu$s probe pulse near $f_\mathrm{res}$ under three conditions: with the switch held at its transmitting bias (blue), held at its isolating bias (orange), and biased isolating but with $\Phi_\mathrm{g}$ pulsed to the transmitting value for 500~ns during the probe (green).  The pulsed switch reaches full transmissions within 5~ns and remains flat, $1.2$~dB below the statically transmitting level, which we attribute to residual miscalibration of the pulse amplitude.

To perform qubit readout, the switch must be pulsed from its isolating bias to its transmitting bias for the duration of the readout pulse, changing the impedance environment seen by the resonator. We next verify that doing so degrades neither the readout not the qubit performance.

The switch is integrated with a Lincoln Lab ``standard candle'' package, where fixed-frequency transmon qubits are coupled to a central feedline (device parameters in Appendix \ref{app:qubit}). We measure the qubit temperature and coherences, in two operating modes: static, with the switch held at its transmitting bias throughout the experiment, and pulsed, with the switch biased isolating during the experiment and transmitting only during readout.

We first verify that readout performance is unaffected by pulsing the switch. Readout is performed by driving a $10~\mu$s pulse at $f_\mathrm{res} = 7.171$ GHz. In pulsed operation, the length of the $\Phi_\mathrm{g}$ pulse as shown in Fig.~\ref{fig1}d is extended to pad the readout pulse by 150~ns on both sides, preventing impedance mismatches or reflections of signal photons during measurement. The optimal qubit readout parameters are unchanged when switching between either of the switch operational modes, and the statistical readout fidelity is measured to be 99.07\% (99.82\%) in static (pulsed) operation, see Appendix \ref{app:singleshot} for details. The $\pi$ pulse is calibrated for both switch operational modes, as this qubit is being driven through the feedline, whose boundary condition the switch changes; a dedicated qubit charge line would negate this effect. 

Noise at the qubit frequency creates qubit excitations, while residual photons in modes coupled to the qubit cause dephasing \cite{jin2015thermal, lvov2025thermometry}. We characterize the residual excited state population $p_\mathrm{e}$ and, using a dephasing measurement, the residual resonator photon number \nth. We determine $p_\mathrm{e}$ in two ways: directly from single-shot readout, and from the amplitude contrast of $e$--$f$ Rabi oscillations; these methods are sensitive to different error sources and serve as cross-checks of one another.

In static (pulsed) operation, the single-shot readout gives $p_\mathrm{e} = 1 \pm 3\%$ ($0.5 \pm 0.8\%$), corresponding to an effective qubit temperature of $35 \pm 21$mK ($30 \pm 10$mK). With the same experiment, we also verify that the statistical readout error is unchanged between using the switch in either operating mode, and that no new readout calibration is required, see Appendix \ref{app:singleshot}. Using the $e$--$f$ amplitude oscillation method, we find that in static (pulsed) operation, $p_\mathrm{e}$ is $0.18 \pm 0.11$\% ($0.25 \pm 0.08$\%). This corresponds to an effective qubit temperature of $25.87 \pm 2.36$mK ($27.26 \pm 1.47$mK). Details and raw fits with this method are found in Appendix \ref{app:ef}. For both methods, the two operating modes agree within uncertainty. Both methods indicate that pulsing the switch does not increase the qubit temperature. The qubit coherences are likewise unchanged, with static (pulsed) coherences of $T_1= 139 \pm 33\mu$s ($138 \pm 30\mu$s), and $T_\mathrm{2E}=288 \pm 88\mu$s ($263 \pm 40\mu$s). In both switch operational modes, the $T_\mathrm{2E}$ is within the uncertainty range of the $2T_1$ limit, indicating that pulsing the switch does not add measurable residual resonator photons.

These measurements show that, within our method sensitivity limits, pulsing the switch does not degrade readout, raise the residual \ket{e} state population, or increase residual resonator photon numbers. Neither is improvement expected from the switch in these experiments, as the three 20dB circulators used in the setup already suppress amplifier noise to $\bar{n} \sim 10^{-5}$, see Appendix~\ref{app:qubit}, below our detection floor. The residual qubit population itself, common to both modes, corresponds to $\sim$30~mK, above the measured 15~mK mixing-chamber temperature, and consistent with residual population in other experiments~\cite{jin2015thermal, wang2019cavity}. 

Since our qubit is already protected by the circulators, we determine the switch's noise suppression by injecting broadband white noise on its output port during the qubit's free evolution and measuring the qubit's dephasing rate. The noise is generated by mixing a 50~MHz white-noise source with a carrier tone of power $P_\mathrm{d}$ at frequency $f_\mathrm{d} = \delta_\mathrm{d} + f_\mathrm{res}$ ~\cite{schuster2007resolving, yan2018distinguishing}, as shown in Fig.~\ref{fig2}a. The setup is described and characterized in Appendix~\ref{app:noisegen}. The mixed noise's 100~MHz bandwidth far exceeds the resonator linewidth ($\kappa_\mathrm{res} = 0.11$~MHz), making it a good proxy for resonator frequency thermal noise.

By varying $\delta_\mathrm{d}$ and $P_\mathrm{d}$, we control the number of thermal noise photons \nth ~populating the resonator, similar to the procedures of~\cite{yan2016flux, yan2018distinguishing, wang2019cavity}. As shown in the pulse diagram of Fig.~\ref{fig2}a, the noise is gated on only during the free-evolution window of each coherence experiment, so readout contrast is preserved even as we reach \nth$\sim200$ photons. Example $T_1$ and $T_\mathrm{2E}$ decays with +5dBm of injected noise and without injected noise are shown in Fig.~\ref{fig2}b, c.

\begin{figure}[htbp]
\centering
\includegraphics[width=\columnwidth]{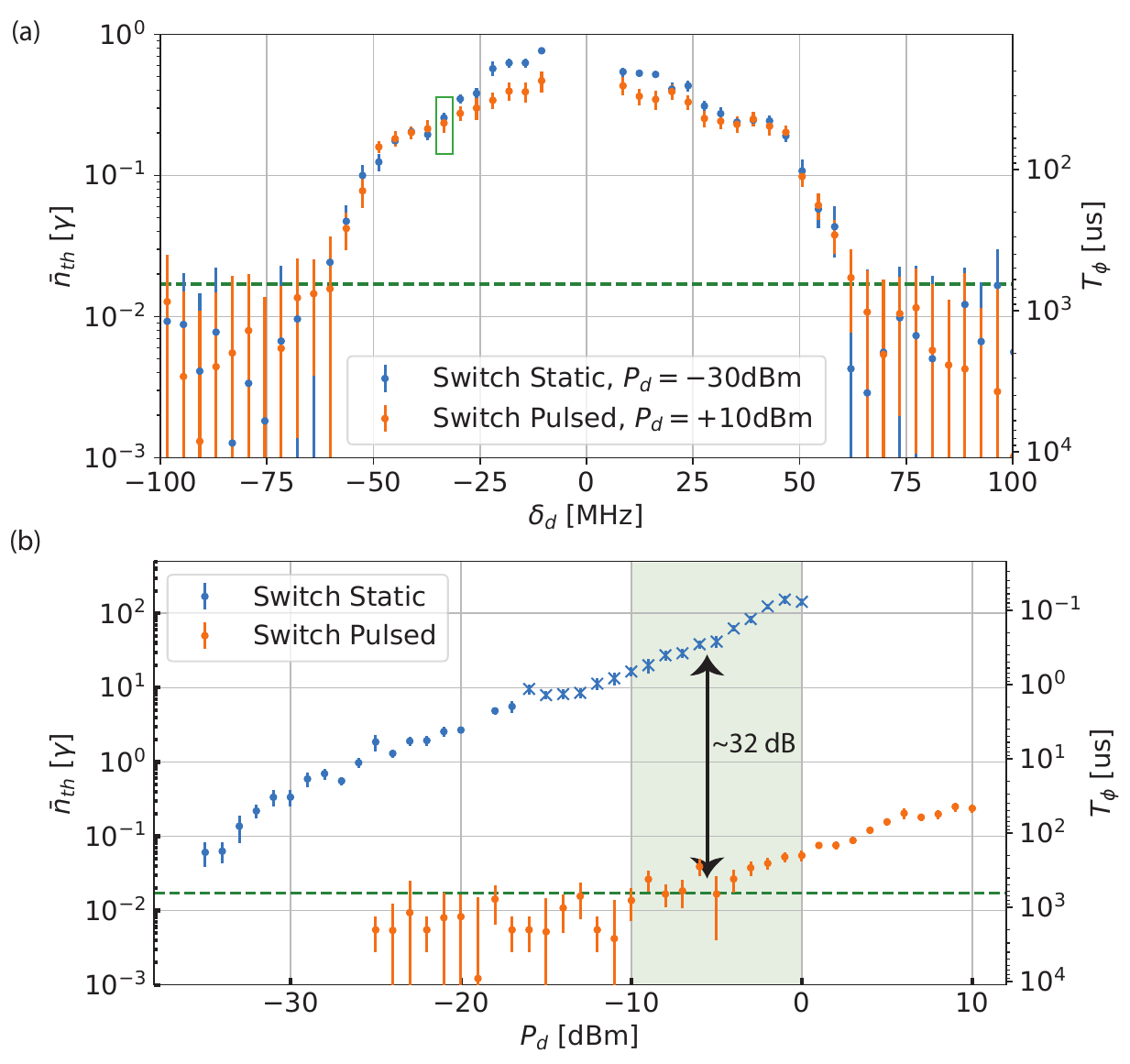}
\caption{
Injected broadband noise simulating readout resonator thermal photons increases the dephasing rate, from which we extract the simulated readout resonator thermal photon population. (a) Sweeping the detuning frequency $\delta_\mathrm{d}$ of the injected noise shows the convolution of the injected noise with the Lorentzian linewidth of the resonator. The blue trace is measured with the switch in static operation and $P_\mathrm{d} = -30dBm$, coinciding with the orange trace, measured with the switch in pulsed operation and an $P_\mathrm{d} = +10dBm$. This demonstrates the $\sim 40$dB protection provided by the switch, also validating that the noise profile is unchanged regardless of $P_\mathrm{d}$. 
(b) The power of the injected noise is swept from -40dBm to +10dBm while $\delta_\mathrm{d}$ is fixed to be $-30$~MHz. The dephasing rate of the qubit is measured and converted into thermal photon number \nth. In blue (orange), the switch is in its static (pulsed) operational mode. Points labeled with a circle use an echo-style coherence measurement, with a refocusing $\pi$ pulse halfway through the wait period; points labeled with an $\times$ use a Ramsey-style experiment. The green shaded region marks the range of injected noise powers for which reliable \nth measurements are found in both the static and pulsed operational modes. Averaging the measured suppression over this range indicates a switch suppression ratio of 32~dB.
} 
\label{fig3}
\end{figure}

For every measured dephasing rate, we compute the resonator photon number 
\begin{equation}
\label{eqn:nth}
    \bar{n}_{tot} = \frac{\kappa_\mathrm{res}^2 + \chi_\mathrm{qr}^2}{\kappa_r \chi_\mathrm{qr}^2} \Gamma_\phi
\end{equation} 

where $\kappa_\mathrm{res} = 0.11$ MHz is the resonator linewidth and $\chi_\mathrm{qr} = 0.240$ MHz is the qubit-resonator dispersive shift. Sweeping the carrier detuning $\delta_\mathrm{d}$ across a 200~MHz neighborhood of $f_\mathrm{res}$ maps the injected spectrum onto the resonator, shown in Fig.~\ref{fig3}a. $\Gamma_\phi$ follows the convolution of the noise spectrum with the resonator lineshape, symmetric about $f_\mathrm{res}$, i.e. $\delta_\mathrm{d} = 0$. For $|\delta_\mathrm{d}| \gtrsim \mathrm{BW}/2$, the dephasing rate is below our detection floor (Appendix~\ref{app:dephasing}).  

Finally, we sweep $P_\mathrm{d}$ from $-40$~dBm to $+10$~dBm and convert measured dephasing rates to readout resonator thermal photon \nth, shown in Fig. \ref{fig3}b. With the switch transmitting during free evolution, \nth\ rises linearly with injected power, when above the detection threshold (see Appendix \ref{app:dephasing}). Above $P_\mathrm{d}= 0$~dBm, residual noise photons corrupt readout enough such that no static-operation data can be taken. With the switch in its pulsed mode, isolating during free evolution, \nth\ remains below the detection floor until $P_\mathrm{d} \approx -10$dBm, at which point photons leaking through the switch become detectable. When injected noise reduces the coherence below 10~$\mu$s, we replace the echo sequence with a Ramsey sequence (points marked with $\times$ in Fig.~\ref{fig3}a). The echo sequence suppresses low-frequency noise at $f\ll 1/t_\mathrm{wait} = $125~kHz, where $t_\mathrm{wait}=8\mu$s is the total evolution time at which we switch over to the Ramsey sequence. Therefore, the resonator thermal photon fluctuation $\kappa=$110~kHz cannot be effectively refocused by the echo sequence in this regime and the two sequence produces the same dephasing rates, as verified in Appendix \ref{app:dephasing}. The green shaded region marks between $P_\mathrm{d} = -10$ to $P_\mathrm{d} = 0$ dBm, where photon number measurements are reliable in both static and pulsed operations. Both datasets vary linearly with injected noise power here, and averaging the ratio of residual \nth~in this region indicates 32dB of suppression, consistent with the directly measured on-off ratio in Fig.~\ref{fig1}c. 

In conclusion, we have integrated a flux-pulsed cryogenic microwave switch, the TIB, into the readout chain of a fixed-frequency transmon. The switch maintains a 25--40~dB on-off ratio across several GHz, actuates within $\sim 5$ nanoseconds of a single baseband flux pulse, and can be pulsed to be transmitting during readout without degrading readout fidelity, residual excited-state population, or qubit coherence. Under injected broadband noise centered at the resonator frequency mimicking amplifier backaction induced resonator thermal photons, the isolating switch suppresses noise-induced dephasing by more than 30~dB, matching its transmission on-off ratio. These results show that this device could replace ferrite-based circulators and isolators in the qubit readout chain, removing their footprint, thermal load, and magnetic shielding requirements. The benefit grows at higher readout frequencies, where thermal occupations are smaller and the switch's isolation can render residual photons entirely negligible~\cite{anferov2024superconducting}. Because switches can be cascaded in series, complex routing networks for multiplexed readout and quantum communication are also possible. While past work has demonstrated the successful integration of the TIB with quantum parametric amplifiers~\cite{rosenthal2021efficient}, a natural next step is to integrate the TIB directly on-chip within the qubit readout package, improving impedance matching, reducing insertion loss, and minimizing the physical footprint.

\begin{acknowledgments}
This work was supported by the Army Research Office under Grant No.~W911NF2310101. We acknowledge support from Q-SEnSE under: NSF QLCI Award 2552714. Qubit chip was fabricated and provided by the Superconducting Qubits at Lincoln Laboratory (SQUILL) Foundry at MIT Lincoln Laboratory, with funding from the Laboratory for Physical Sciences (LPS) Qubit Collaboratory. E. I. R acknowledges support from the ARO QuaCGR fellowship. P.V. acknowledges support by the National Science Foundation Graduate Research Fellowship Program under Grant No. DGE-2146755. Any opinions, findings, and conclusions or recommendations expressed in this material are those of the authors and do not necessarily reflect the views of the National Science Foundation.
\end{acknowledgments}

\section*{Author Declarations}

\section*{Conflict of Interest}

The authors have no conflicts to disclose.

\section*{Data Availability Statement}

The data that support the findings of this study are available from the corresponding author upon reasonable request.

\section*{Author Contributions}

\textbf{Chunyang Ding:} Conceptualization (lead); Data curation (lead); Formal analysis (lead); Investigation (lead); Methodology (lead); Software (lead); Visualization (lead); Writing – original draft (lead); Writing – review \& editing (lead).
\textbf{Chuyao Tong:} Conceptualization (equal); Formal analysis (equal); Investigation (equal); Methodology (equal); Software (equal); Writing - review \& editing (equal).
\textbf{Paul Varosy:} Investigation (supporting); Methodology (supporting); Software (equal); Writing - review \& editing (equal).
\textbf{Nikitha Jain:} Investigation (supporting); Methodology (supporting); Writing - review \& editing (equal).
\textbf{Ziyi Zhao:} Conceptualization (supporting); Resources (equal); Writing - review \& editing (supporting).
\textbf{Eric I. Rosenthal:} Conceptualization (supporting); Resources (equal); Writing - review \& editing (supporting).
\textbf{Konrad W. Lehnert:} Conceptualization (equal); Funding acquisition (equal); Project administration (equal); Supervision (equal); Writing - review \& editing (equal)
\textbf{David I. Schuster:} Conceptualization (equal); Funding acquisition (equal); Project administration (equal); Supervision (equal); Writing - review \& editing (equal)

\appendix 

\section{Qubit characterization and fridge thermalization}
\label{app:qubit}

\begin{figure}[htbp]
\centering
\includegraphics[width=\columnwidth]{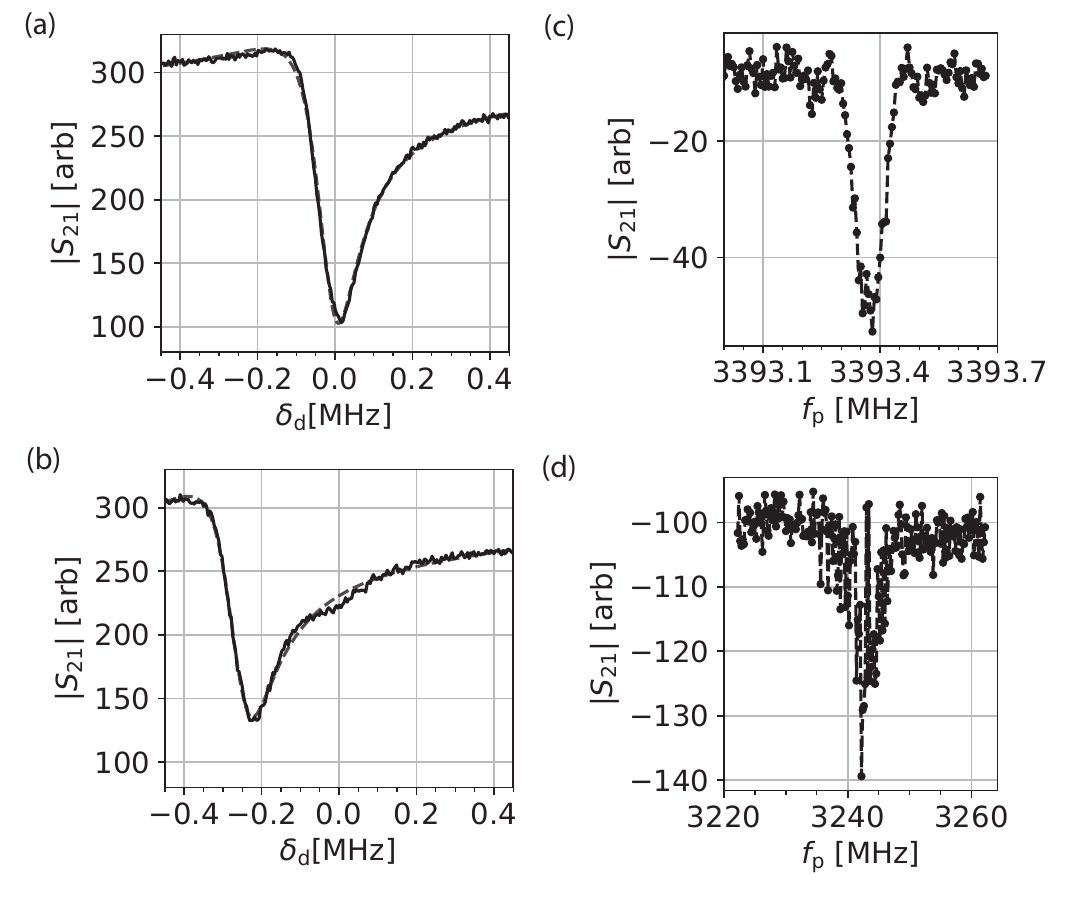}
\caption{Characterization plots of the qubit and resonator. (a, b) Resonator spectroscopy with the qubit in it's \ket{g} and \ket{e} states, respectively. Note that $\delta_\mathrm{d}$ is chosen such that when $\delta_\mathrm{d} = 0$, $f_\mathrm{d} = 7.1717$~GHz. The measured $\chi_\mathrm{qr} = 0.240$~MHz. (c, d) Qubit spectroscopy of the $g$--$e$ transition and the $e$--$f$  transition, respectively.}
\label{fig:ResChar}
\end{figure}

The Lincoln Labs' ``Standard Candle'' consists of a single central feedline and six CPW $\lambda$/2 resonators in a ``hanger'' configuration. We measure only one of these fixed-frequency transmons for these experiments, although the TIB could be used for simultaneous readout of all six of these qubits. This transmon has parameters of $f_\mathrm{res} = 7.1717$GHz, $f_\mathrm{ge} = 3.393$ GHz, $f_\mathrm{ef} = 3.242$ GHz. The resonator has parameters of $\kappa_\mathrm{res} = 0.11$MHz, $\chi_\mathrm{qr} = 0.240$ MHz. See Fig.~\ref{fig:ResChar} for raw data.

The sample was loaded in a $\mu$-metal container with IR filters (LNF QMC-003) on the input and output. The experiment was performed in a Bluefors LD50 fridge. There are 70dB of attenuators on the input line, thermalized at each stage of the fridge, and 60dB of attenuation prior to the -6dB directional coupler (Krytar 102020006) carrying the noise injection line. The noise injection line is connected to the third port of the first circulator (LNF-DJ4\_8) after the TIB, permitting the noise to bypass the circulators. See Appendix \ref{app:noisegen} for details on the noise injection. 

\begin{figure}[htbp]
\centering
\includegraphics[width=\columnwidth]{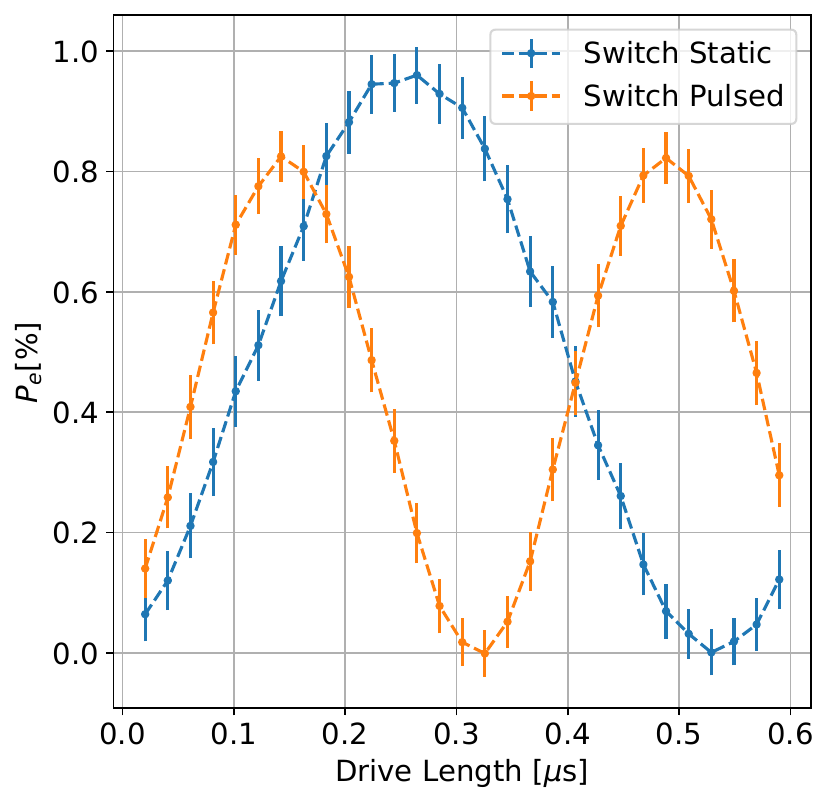}
\caption{$g$--$e$ Rabi oscillations of the qubit, measured under different switch operational modes. The y-axis of $P_\mathrm{e}$ is the qubit excited state population, which is normalized using the single shot readout measurements. Error bars on the point indicate one standard deviation of uncertainty for the 10,000 averaged repetitions.}
\label{fig:RabiDiffs}
\end{figure}

There are no dedicated qubit charge-drive lines, so the charge drive line for enabling single-qubit gates is driven through the central feedline, as indicated on Fig.~\ref{fig1}. This meant that different pulse calibrations for when the switch was transmitting or reflecting, as the switch's state changes a transmitting boundary into a reflecting boundary. Effectively, when the switch was reflecting, the calibrated $g$--$e$ pulse has half the gate duration compared to when the switch was open, indicating that the rf pulse reflects off the boundary created by the switch and interacts with the qubit twice, see Fig.~\ref{fig:RabiDiffs}. As we note in the main text, this effect comes primarily because there is no dedicated qubit drive line; including that drive line would permit a single calibrated pulse to be used regardless of the operational mode of the switch. 

We compute the qubit temperature in both static and pulsed operation using the Bose-Einstein distribution. The measured qubit temperature of $\sim 30$~mK during both switch operational modes is within a factor of two compared to the RuOx thermometer measured temperature of the mixing chamber in our dilution refrigerator. We also calculate the expected number of residual photons due to the temperature of the stages and the in-line isolation. Our dilution refrigerator has a measured $2.9$~K temperature at the HEMTs and a $15$~mK base temperature. Using a Bose-Einstein distribution, the temperature at the HEMTs corresponds to $\sim8 f_\mathrm{res}$ photons. Since we use three circulators in series on the output line, we suppress photons from this stage by approximately 60dB, such that \nth $= 8 \times 10^{-6}$ resonator photons arrive at the qubit output. This is approximately the same order of magnitude of residual resonator photons that we would expect given a 30mK base temperature, which is calculated to be \nth $= 1.05 \times 10^{-5}$ photons. Thus, we do not expect to see a significant difference in qubit performance with the switch biased transmitting versus isolating, as the number of blocked photons is below our resolution limit. 

\section{Readout Characterization} 
\label{app:singleshot}

\begin{figure}[ht]
\centering
\includegraphics[width=\columnwidth]{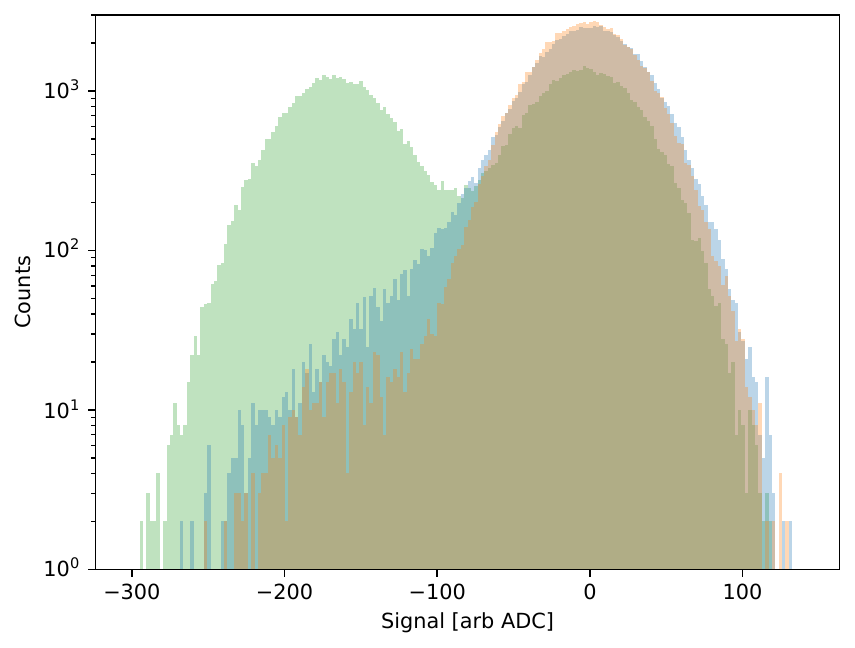}
\caption{Single shot readout performance of the qubit, with the switch in both its static and pulsed operation modes. In green, the qubit is initialized using a $\sigma_{x,\pi/2}$ pulse, preparing the \ket{+} state. In blue (orange), the qubit is initialized in its \ket{g} state and the switch is in its static (pulsed) operational mode.}
\label{fig:singleshot}
\end{figure}

To quantify the readout performance, we first prepare the state $|+\rangle = (1/\sqrt{2}) (|g\rangle + |e\rangle)$ using a $\sigma_{x, \pi/2}$ charge drive pulse, and perform dispersive readout, measuring the resonator displacement through single shots. Collapsed onto the maximal contrast axis, we fit the histogram distribution to a double Gaussian function. We repeat this process for when the switch is in its static operational mode, as well as when the switch is pulsed. The statistical readout fidelity, as the overlap percentage of the fitted Gaussian envelopes, is found to be 99.07\% for the static operation of the switch, and 99.82\% for the pulsed operation of the switch. This is shown as the green histogram in Fig.~\ref{fig:singleshot}. 

We then initialize the qubit in its ground state by waiting $\geq 10 T_1$ between each measurement. We use the previously fitted $\mu, \sigma$ of the double Gaussian to constrain the new fit, finding the new amplitudes of the qubit ground and excited Gaussian distributions. We use these amplitudes to compute the enclosed area, and using the ratio of areas for the qubit ground and excited state populations, we compute the residual \ket{e} population, $P_e$, measuring both in the static and pulsed operational modes of the switch. As this method requires no additional active qubit prepations steps, we avoid any errors that might arise from preparing the qubit with different switch operational modes, quantifying the switch's effect on readout alone. 

Uncertainties of $p_\mathrm{e}$ from using this method include measurement discrimination infidelity, fit uncertainty of the amplitudes in the Gaussian fit, and the heating of the qubit during the relatively long ($10 \mu$ s) readout pulse. We determine that the dominant source of uncertainty is in the uncertainty of the fit amplitude. Therefore, we compute the uncertainty of the final $p_\mathrm{e}$ by first propagating the fit uncertainty into the area calculation, and subsequently propagating in computing the ratio and $p_\mathrm{e}$. 

\section{$e$--$f$ Rabi}
\label{app:ef}

\begin{figure}[htbp]
\centering
\includegraphics[width=\columnwidth]{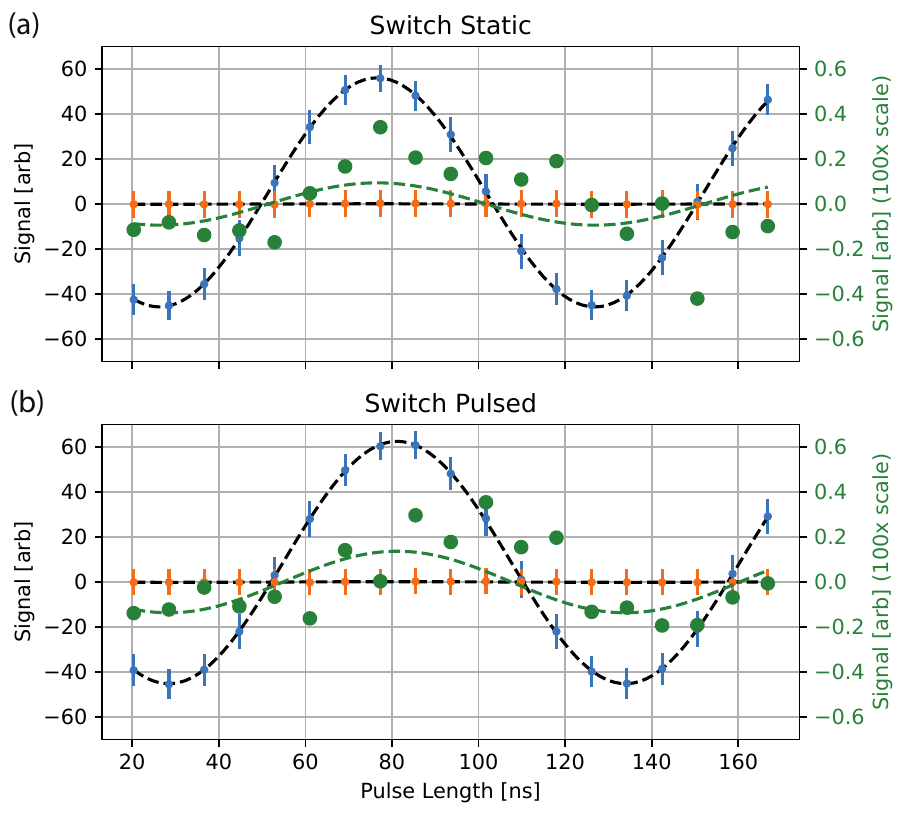}
\caption{$e$--$f$ Rabi oscillations, with the switch in its static and pulsed operational modes. The black dashed line is the best fit. Blue dots indicate the oscillation when the excited state is prepared using a $\pi_\mathrm{ge}$ pulse, while orange dots indicate the Rabi measurement where the qubit begins from a thermal distribution. We replot the same thermal distribution data with an axis that is 100 times zoomed in using the green dots for visibility. (a) indicates the Rabi oscillation with the switch in its static operation, and (b) is the Rabi oscillation in pulsed operation.}
\label{fig:EFRabi}
\end{figure}

Since $P_e$ is small ($\lessapprox 1\%$) and because the single-shot measurement of estimating $P_e$ has higher uncertainties ($\approx 1\%$), we use a higher-precision method, adapted from \cite{geerlings2013demonstrating, jin2015thermal}. We drive a $e$--$f$ Rabi oscillation, which has an amplitude that is proportional to the \ket{e} state population. By actively preparing \ket{e} on our qubit and comparing its oscillation amplitude (the reference amplitude, $A_\mathrm{r}$) to the oscillation of the same frequency seen in the thermal residual population of \ket{e} (the signal amplitude, $A_\mathrm{s}$), we can determine $P_e$ with higher accuracy. Since $A_\mathrm{s}$ is expected to be small, we perform 10 times the number of repetitions when performing the signal Rabi experiment, compared to the reference experiment, see Fig.~\ref{fig:EFRabi}. As the amplitude of the reference and signal EF Rabi oscillations are equally affected by measurement infidelity, taking the ratio of the two amplitudes, $R = A_s/A_r$, allows us to determine $P_{\mathrm{e, th}}$ independently of potential measurement errors, although it is still sensitive to the preparation fidelity of the reference \ket{e} state. 

When performing the fit of the data to find $A_\mathrm{s}, A_\mathrm{r}$, we first do a free fit of amplitude, frequency, phase, and offset on the oscillation seen using the reference state, which has a large signal-to-noise and a clear contrast. Using those fit parameters, we then constrain the model and fit only the amplitude of the signal oscillation.

To determine the uncertainty of this metric with an imperfect \ket{e} preparation pulse, suppose that the qubit is initialized in a thermal distribution, such that its initial distribution can be described as a statistical mixture with probabilities of $P_{\mathrm{e, th}}$ for the $|e \rangle $ state and $ (1 - P_{\mathrm{e, th}})$ for the $|g\rangle$ state.

Then, if we have a $\pi_{ge}$ pulse calibrated with a gate fidelity of $F$. applying this $\pi_{ge}$ pulse prepares \ket{e} with the probability of 

\begin{equation}
\begin{split}
    P'_e &=  F \cdot (1- P_{\mathrm{e, th}}) + (1-F) \cdot  P_{\mathrm{e, th}} \\
    &= F +  P_{\mathrm{e, th}} - 2 F  P_{\mathrm{e, th}}
\end{split}
\end{equation}

Next, the amplitude of the $e$--$f$ Rabi oscillations is proportional to the population in each state, such that 

$$ A_s = X P_{\mathrm{e, th}}; A_r = X P_e'$$

where $X$ is a factor that takes into account the readout infidelity. By determining the ratio of the oscillations, $X$ is cancelled out:

\begin{equation}
    R = \frac{A_s}{A_r} = \frac{P_{\mathrm{e, th}}}{F + P_{\mathrm{e, th}} - 2 F P_{\mathrm{e, th}}}
\end{equation}

which simply reduces to 

\begin{equation}
    \label{eqn:ef}
    P_{\mathrm{e, th}} = \frac{F}{ 1/R - 1 + 2F}
\end{equation}

We measure the $\pi_{ge}$ gate fidelity $F$ as 94.4\%, and combined with the measured ratio $R$, determine $P_{\mathrm{e, th}}$ along with its uncertainty for each of the methods for measuring excited state population to be $0.18 \pm 0.11$\% when the switch is in its static operation mode, and $0.25 \pm 0.08$\% when the switch is in its pulsed operational mode. There is no significant statistical difference between these measured populations, so we conclude that pulsing the switch does not creating excess heating, induce extra noise, or create residual resonator photons that prevent readout. 

\section{Noise Generation}
\label{app:noisegen}

\begin{figure}[htbp]
\centering
\includegraphics[width=\columnwidth]{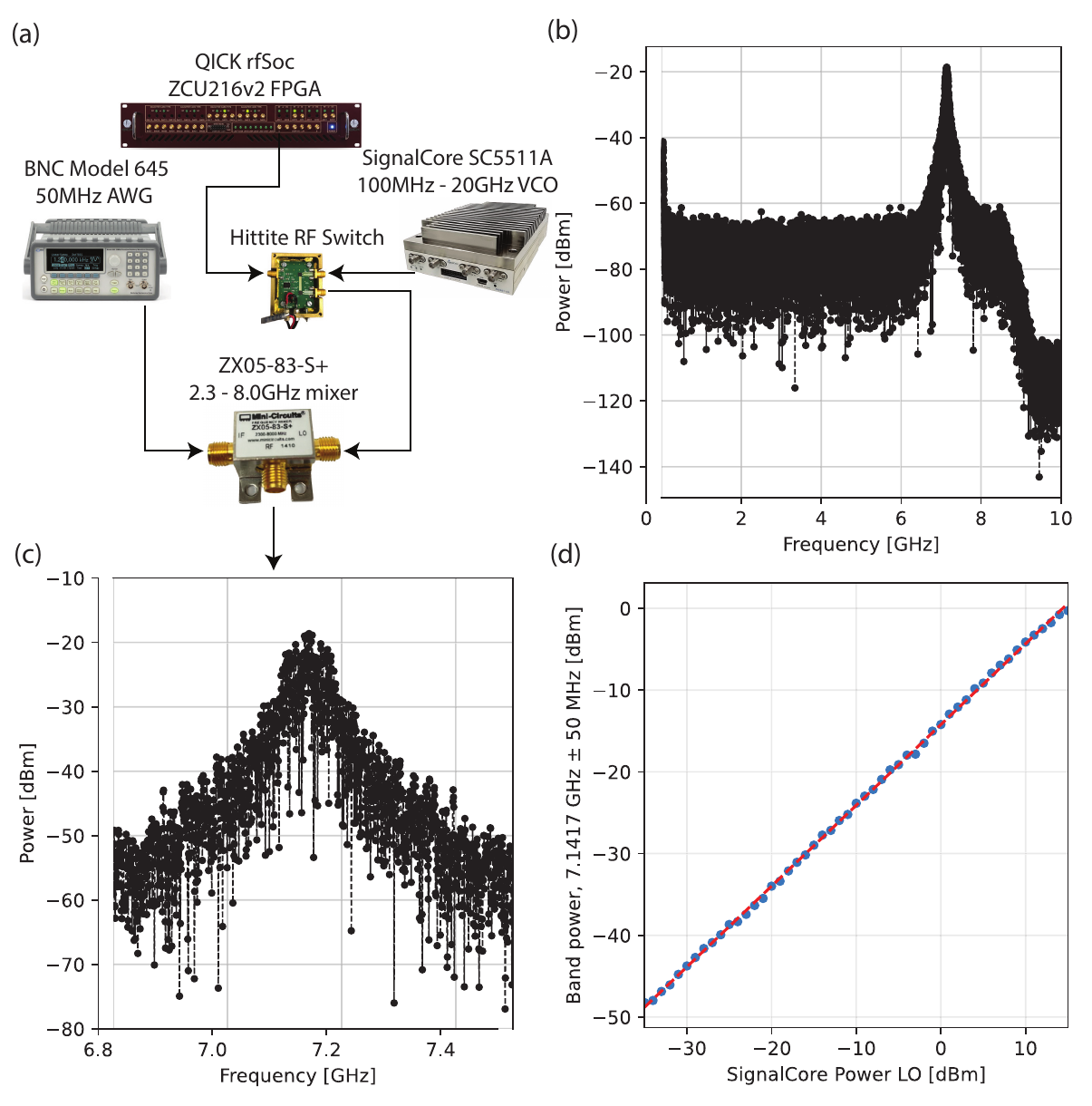}
\caption{(a) Schematic of noise generation (b) Noise spectrum, using a 2$\mu$s window time trace on a sampling oscilloscope, generated with $P_\mathrm{d} = +10$~dBm and $f_\mathrm{d} = 7.1417$~GHz. (c) Close zoom-in of the FFT, in the window near $f_\mathrm{d}$. (d) Variation of the average measured power density in the 100~MHz window around $f_\mathrm{d}$ }
\label{fig:NoiseGen}
\end{figure}

To create the 100~MHz bandwidth white noise near the resonant frequency of our cavity, we mix white noise, sourced from a 50~MHz BNC AWG (Model 645), on the IF port with a carrier tone, sourced from a SignalCore VCO (Model SC5511A) on the LO port. The microwave mixer is a MiniCircuits ZX05-83-S+ and is rated for 2.4 -- 8.0~GHz. See Figure \ref{fig:NoiseGen}(a) for a connectivity diagram. 

Using these instruments, we generate constant white noise from the AWG, but are able to control the output noise power and frequency by tuning those respective parameters on the LO, using the SignalCore VCO. The SignalCore VCO is tunable across a $P_\mathrm{d}$ range of -35 to +10~dBm while maintaining linearity, and can operate with $f_\mathrm{d}$ between 100~MHz and 20~GHz. 

To control the noise injection process, we trigger noise pulses using a TTL controlled room-temperature RF switch. This RF switch gates the output of the SignalCore RF generator that provides the carrier tone, effectively turning on and off the output from the frequency-converted noise pulse. The TTL signal is controlled using the QICK rfSoc FPGA with the RF Attachment Board that has integrated DC and TTL control outputs,~\cite{stefanazzi2022qick, ding2024experimental}, where the same board also controls the microwave pulses and digitization ADCs. This ensures nanosecond-level synchronization of the noise triggering with the microwave pulses used for performing readout of the qubit. 

As shown in Fig.~\ref{fig1}(a), there is 60~dB of attenuation between the room-temperature input of the noise injection line to the mixing-chamber mounted directional coupler. The noise line is then coupled to the output of the TIB via a -6dB directional coupler (Krytar 102020006). This provides sufficient protection against residual thermal photons from 4K and above, ensuring that our qubit is well thermalized and that the residual excited state population is consistent with a 30~mK base temperature, despite the presence of this additional line. 

We verify that the generated noise has the desired characteristics. To do so, we measure the output of the mixed noise on a digital oscilloscope (Keysight Infiniium S-Series DSOS804A) with a 8GHz analog bandwidth and a 20GSa/second sampling rate. With the LO fixed at $f_\mathrm{d} = 7.1417$~GHz, the same frequency used in the injected noise experiment of Fig 3(b), recording 2~$\mu$s of data. We perform a Fourier transform analysis, with a sample spectrum where $P_\mathrm{d} = +10$~dBm is shown in Fig. \ref{fig:NoiseGen}(b). Note that the rolloff past 8GHz corresponds to the analog bandwidth of our oscilloscope, and the low frequency peak is likely IF leakage. In Fig \ref{fig:NoiseGen}(c), we show the same data within a 600~MHz window near $f_\mathrm{d}$ to illustrate the noise profile. 

Finally, to ensure that the noise output is linear with respect to the LO $P_\mathrm{d}$ input, we sweep $P_\mathrm{d}$ from -35 to +10~dBm. The average power in the $100$~MHz bandwidth window around $f_\mathrm{d}$ is computed and plotted relative to the $P_\mathrm{d}$, shown in Fig.~\ref{fig:NoiseGen}(d). We find a strong linear relationship in this region, with a slope of 0.99 and a $R^2$ coefficient of 0.9996. 

\section{Dephasing time}
\label{app:dephasing}

Residual photons at the resonator frequency effectively preform weak perturbative measurements on the qubit, thus leading to dephasing of the qubit. Therefore, a measurement of the $T_\phi$ coherence time, as determined by the $T_{\mathrm{2E}}$ and $T_\mathrm{1}$ coherence times, would quantify the number of residual resonator photons. From the dephasing rate $\Gamma_\phi$, the resonator frequency $f_\mathrm{res}$, and linewidth $\kappa_\mathrm{res}$, we can determine the number of residual resonator photons $\bar{n}_\mathrm{th, residual}$. 

\begin{figure}[htbp]
\centering
\includegraphics[width=\columnwidth]{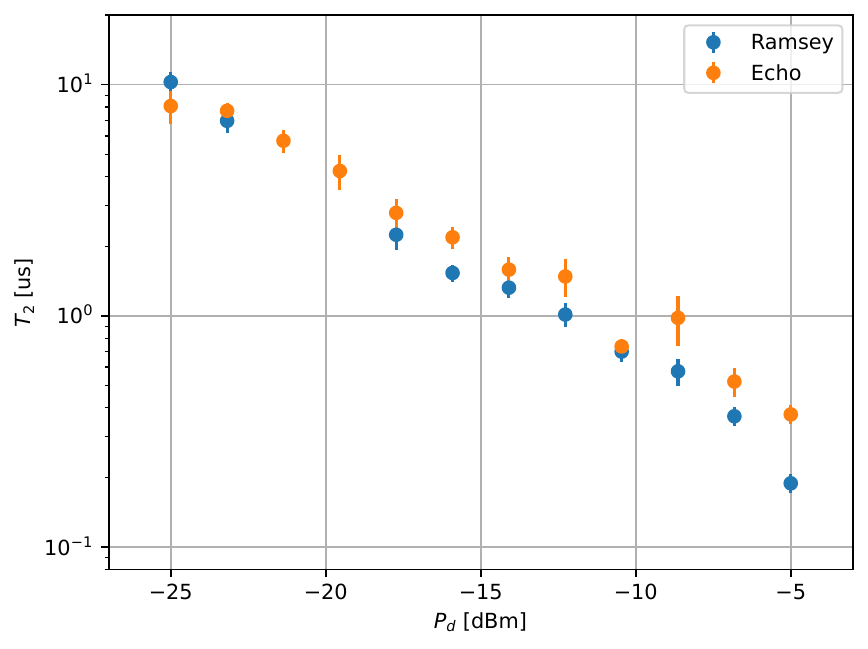}
\caption{Performing both Ramsey-style and echo-style coherence measurements with the switch in it's transmitting position. Orange points indicate an echo-style experiment, where a $\pi_\mathrm{x}$ pulse refocuses the qubit's evolution, while blue points omit that additional pulse. Note that in this range of measured $T_2 \leq 10\mu$s, the two different experiment styles report similar coherence times, indicating that the additional $\pi_\mathrm{x}$ pulse does not make a significant difference in measured coherence time.}
\label{fig:RamseyEcho}
\end{figure}

Using these measured coherences, we compute the dephasing rate as 

\begin{equation}
    \frac{1}{T_\phi} = \frac{1}{T_{2E}} - \frac{1}{2T_1} = \Gamma_\phi
\end{equation}

Given the finite $T_1$ and its fluctuations over time, there is a bound to the maximum resolvable dephasing time that can be measured. Suppose that we measure the relaxation coherence time $ T_1 = A \pm \delta_A$, and that at its limit, the echoed dephasing time $ T_{2E} \leq 2A \pm 2\delta_A$. Then, we can compute each individual dephasing rates as 

\begin{equation}
    \Gamma_1 = \frac{1}{T_1} =  \frac{1}{A} \pm \left( \frac{\delta_A}{A} \cdot \frac{1}{A} \right) = \frac{1}{A} \pm \frac{\delta_A}{A^2}
\end{equation}

and 

\begin{equation}
    \Gamma_2 = \frac{1}{T_{2E}} \geq \frac{1}{2A} \pm \left( \frac{2\delta_A}{2A} \cdot \frac{1}{2A} \right) = \frac{1}{2A} \pm \frac{\delta_A}{2A^2}
\end{equation}

Then, if we compute $\Gamma_\phi$ using these individual decoherence rates, we find 

\begin{equation}
    \Gamma_\phi = \Gamma_2 - 0.5 \Gamma_1 \geq \left(\frac{1}{2A} \pm \frac{\delta_A}{2A^2} \right) - \left( \frac{1}{2A} \pm \frac{\delta_A}{2A^2} \right)
\end{equation}

Such that 

$$\Delta_{\Gamma_\phi} \geq \delta_A/A^2$$

Although the percent error for $\Gamma_\phi$ is undefined, we use the absolute uncertainty computed here as the noise floor of this measurement, as we can not be confident about a $\Gamma_\phi$ lower than this value given the time-domain instability of these measurements. From our time-averaged coherence measurements, where we find $T_1 = 138 \pm 32 \mu$s, we can conclude that the minimum $\Gamma_\phi$ that can be measured is equal to $1.6 \times 10^{-3}$ MHz. Given our resonator $\kappa$ and $\chi$ parameters, this is equivalent to a minimum \nth $= 1.7 \times 10^{-2}$ photons. Using the Bose-Einstein distribution, this means that the minimum temperature that we are sensitive to is 

\begin{equation}
\label{eq:boseEinstein}
    T \geq \frac{hf}{k \log (1 + 1/ \bar{n}_{min} )} = 84\mathrm{mK}
\end{equation}

When performing the coherence experiment with different $P_d$ for the injected noise, we vary the total delay time based on the approximate magnitude of the injected noise and an initial rough calibration of expected coherence times. This is primarily because when injecting large amounts of noise, the coherence time changes by 3-4 orders of magnitude. Thus, a constant total delay time is impractical for accurately measuring the qubit decoherence across the entire range of injected resonator thermal photons. The total delay time for each experiment is at least 1.5 times larger than the measured decay value at every point, and is never less than 10$\mu$~s. 

In the presence of many resonator thermal photons, the refocusing $\pi_\mathrm{x}$ pulse in an echo-style experiment becomes ineffective, as the qubit's dephasing rate is high, and there is also no need to be protecting against low frequency noise. Therefore, when the total coherence time of the qubit is less than 10$~\mu$s, we switch from measuring the coherence using an echo-style experiment, where a refocusing $\pi_\mathrm{x}$ pulse is played halfway through the delay time, to using a pure Ramsey-style experiment. We verify that doing so does not meaningfully change the measured decoherence rates by performing both experiments at relatively low injected noise photon levels, as shown in Fig.~\ref{fig:RamseyEcho}.  

We do recognize that the limited total delay time of the coherence experiments may have been a limitation in accurately measuring the qubit coherence time, and is the primary effect of the measurement uncertainty, rather than time-dependent qubit coherence fluctuations. To quantify this uncertainty, we perform a Monte Carlo simulation of simulated decoherence data with realistic measurement noise in the qubit state population, while using a least-squares fit to extract the fit parameters. 

Overall, we note that our experimental delay time was overestimating the qubit coherence time of the qubit by approximately 25\%. However, we do note that the primary results here regarding the on-off ratio of the switch is not affected by this discrepancy, as this overestimation applies equally to both coherence measurements. This underestimation should be consistent in the measurement of the qubit dephasing rate both with the switch in its reflecting and in its transmitting states. Therefore, the on-off ratio reported here ought to be consistent.

\bibliography{refs}

@article{jin2015thermal,
  title={Thermal and residual excited-state population in a 3D transmon qubit},
  author={Jin, XY and Kamal, A and Sears, AP and Gudmundsen, T and Hover, D and Miloshi, J and Slattery, R and Yan, F and Yoder, J and Orlando, TP and others},
  journal={Physical Review Letters},
  volume={114},
  number={24},
  pages={240501},
  year={2015},
  publisher={APS}
}

@article{geerlings2013demonstrating,
  title={Demonstrating a driven reset protocol for a superconducting qubit},
  author={Geerlings, Kurtis and Leghtas, Zaki and Pop, Ioan M and Shankar, Shyam and Frunzio, Luigi and Schoelkopf, Robert J and Mirrahimi, Mazyar and Devoret, Michel H},
  journal={Physical review letters},
  volume={110},
  number={12},
  pages={120501},
  year={2013},
  publisher={APS}
}

@article{zhao2023integrating,
  title={Integrating planar circuits with superconducting 3D microwave cavities using tunable low-loss couplers},
  author={Zhao, Ziyi and Gurra, Eva and Rosenthal, Eric I and Vale, Leila R and Hilton, Gene C and Lehnert, KW},
  journal={Applied Physics Letters},
  volume={123},
  number={1},
  year={2023},
  publisher={AIP Publishing}
}

@article{rosenthal2021efficient,
  title={Efficient and low-backaction quantum measurement using a chip-scale detector},
  author={Rosenthal, Eric I and Schneider, Christian MF and Malnou, Maxime and Zhao, Ziyi and Leditzky, Felix and Chapman, Benjamin J and Wustmann, Waltraut and Ma, Xizheng and Palken, Daniel A and Zanner, Maximilian F and others},
  journal={Physical review letters},
  volume={126},
  number={9},
  pages={090503},
  year={2021},
  publisher={APS}
}

@article{wang2019cavity,
  title={Cavity attenuators for superconducting qubits},
  author={Wang, Z and Shankar, S and Minev, ZK and Campagne-Ibarcq, Philippe and Narla, A and Devoret, Michel H},
  journal={Physical Review Applied},
  volume={11},
  number={1},
  pages={014031},
  year={2019},
  publisher={APS}
}

@article{yan2016flux,
  title={The flux qubit revisited to enhance coherence and reproducibility},
  author={Yan, Fei and Gustavsson, Simon and Kamal, Archana and Birenbaum, Jeffrey and Sears, Adam P and Hover, David and Gudmundsen, Ted J and Rosenberg, Danna and Samach, Gabriel and Weber, Steven and others},
  journal={Nature communications},
  volume={7},
  number={1},
  pages={12964},
  year={2016},
  publisher={Nature Publishing Group UK London}
}

@article{yan2018distinguishing,
  title={Distinguishing coherent and thermal photon noise in a circuit quantum electrodynamical system},
  author={Yan, Fei and Campbell, Dan and Krantz, Philip and Kjaergaard, Morten and Kim, David and Yoder, Jonilyn L and Hover, David and Sears, Adam and Kerman, Andrew J and Orlando, Terry P and others},
  journal={Physical Review Letters},
  volume={120},
  number={26},
  pages={260504},
  year={2018},
  publisher={APS}
}

@article{schuster2007resolving,
  title={Resolving photon number states in a superconducting circuit},
  author={Schuster, David I and Houck, Andrew Addison and Schreier, JA and Wallraff, Andreas and Gambetta, Jay M and Blais, Alexandre and Frunzio, Luigi and Majer, J and Johnson, B and Devoret, Michel H and others},
  journal={Nature},
  volume={445},
  number={7127},
  pages={515--518},
  year={2007},
  publisher={Nature Publishing Group UK London}
}

@article{sunada2024photon,
  title={Photon-noise-tolerant dispersive readout of a superconducting qubit using a nonlinear Purcell filter},
  author={Sunada, Yoshiki and Yuki, Kenshi and Wang, Zhiling and Miyamura, Takeaki and Ilves, Jesper and Matsuura, Kohei and Spring, Peter A and Tamate, Shuhei and Kono, Shingo and Nakamura, Yasunobu},
  journal={Prx Quantum},
  volume={5},
  number={1},
  pages={010307},
  year={2024},
  publisher={APS}
}

@article{lvov2025thermometry,
  title={Thermometry based on a superconducting qubit},
  author={Lvov, Dmitrii S and Lemziakov, Sergei A and Ankerhold, Elias and Peltonen, Joonas T and Pekola, Jukka P},
  journal={Phys. Rev. Applied},
  volume={23},
  number={5},
  pages={054079},
  year={2025},
  publisher={APS}
}

@article{bertet2005dephasing,
  title={Dephasing of a flux-qubit coupled to a harmonic oscillator},
  author={Bertet, P and Chiorescu, I and Harmans, CJPM and Mooij, JE},
  journal={arXiv cond-mat/0507290},
  year={2005}
}

@article{chapman2016general,
  title={General purpose multiplexing device for cryogenic microwave systems},
  author={Chapman, Benjamin J and Moores, Bradley A and Rosenthal, Eric I and Kerckhoff, Joseph and Lehnert, KW},
  journal={Applied Physics Letters},
  volume={108},
  number={22},
  year={2016},
  publisher={AIP Publishing}
}

@article{chapman2017single,
  title={Single-sideband modulator for frequency domain multiplexing of superconducting qubit readout},
  author={Chapman, Benjamin J and Rosenthal, Eric I and Kerckhoff, Joseph and Vale, Leila R and Hilton, Gene C and Lehnert, KW},
  journal={Applied Physics Letters},
  volume={110},
  number={16},
  year={2017},
  publisher={AIP Publishing}
}

@article{gambetta2006qubit,
  title={Qubit-photon interactions in a cavity: Measurement-induced dephasing and number splitting},
  author={Gambetta, Jay and Blais, Alexandre and Schuster, David I and Wallraff, Andreas and Frunzio, Luigi and Majer, J and Devoret, Michel H and Girvin, Steven M and Schoelkopf, Robert J},
  journal={Physical Review A—Atomic, Molecular, and Optical Physics},
  volume={74},
  number={4},
  pages={042318},
  year={2006},
  publisher={APS}
}

@article{chen2012multiplexed,
  title={Multiplexed dispersive readout of superconducting phase qubits},
  author={Chen, Yu and Sank, D and O'Malley, P and White, T and Barends, R and Chiaro, B and Kelly, J and Lucero, E and Mariantoni, M and Megrant, A and others},
  journal={Applied Physics Letters},
  volume={101},
  number={18},
  year={2012},
  publisher={AIP Publishing}
}

@article{spring2025fast,
  title={Fast multiplexed superconducting-qubit readout with intrinsic purcell filtering using a multiconductor transmission line},
  author={Spring, Peter A and Milanovic, Luka and Sunada, Yoshiki and Wang, Shiyu and Van Loo, Arjan F and Tamate, Shuhei and Nakamura, Yasunobu},
  journal={PRX Quantum},
  volume={6},
  number={2},
  pages={020345},
  year={2025},
  publisher={APS}
}

@article{zobrist2026measurement,
  title={Measurement-Induced State transitions in Inductively-Shunted Transmons},
  author={Zobrist, Nicholas and Kreikebaum, John Mark and Khezri, Mostafa and Isakov, Sergei V and Lester, Brian J and Zhang, Yaxing and Di Paolo, Agustin and Sank, Daniel and Smith, W Clarke},
  journal={arXiv preprint arXiv:2603.12114},
  year={2026}
}

@article{reed2010fast,
  title={Fast reset and suppressing spontaneous emission of a superconducting qubit},
  author={Reed, Matthew D and Johnson, Blake R and Houck, Andrew A and DiCarlo, Leonardo and Chow, Jerry M and Schuster, David I and Frunzio, Luigi and Schoelkopf, Robert J},
  journal={Applied Physics Letters},
  volume={96},
  number={20},
  year={2010},
  publisher={AIP Publishing}
}

@article{kurilovich2025high,
  title={High-frequency readout free from transmon multi-excitation resonances},
  author={Kurilovich, Pavel D and Connolly, Thomas and B{\o}ttcher, Charlotte GL and Weiss, Daniel K and Hazra, Sumeru and Joshi, Vidul R and Ding, Andy Z and Nho, Heekun and Diamond, Spencer and Kurilovich, Vladislav D and others},
  journal={arXiv preprint arXiv:2501.09161},
  year={2025}
}

@article{chang2020fast,
  title={A fast and large bandwidth superconducting variable coupler},
  author={Chang, H-S and Satzinger, Kevin J and Zhong, YP and Bienfait, Audrey and Chou, M-H and Conner, Christopher R and Dumur, {\'E}tienne and Grebel, Joel and Peairs, Gregory A and Povey, Rhys G and others},
  journal={Applied Physics Letters},
  volume={117},
  number={24},
  year={2020},
  publisher={AIP Publishing}
}

@article{google2025quantum,
  title={Quantum error correction below the surface code threshold},
  author={Google Quantum},
  journal={Nature},
  volume={638},
  number={8052},
  pages={920--926},
  year={2025},
  publisher={Nature Publishing Group UK London}
}

@article{pechal2016superconducting,
  title={Superconducting switch for fast on-chip routing of quantum microwave fields},
  author={Pechal, M and Besse, J-C and Mondal, Mintu and Oppliger, M and Gasparinetti, S and Wallraff, A},
  journal={Physical Review Applied},
  volume={6},
  number={2},
  pages={024009},
  year={2016},
  publisher={APS}
}

@article{thorbeck2017reverse,
  title={Reverse isolation and backaction of the SLUG microwave amplifier},
  author={Thorbeck, T and Zhu, Shaojiang and Leonard Jr, E and Barends, R and Kelly, J and Martinis, John M and McDermott, R},
  journal={Physical Review Applied},
  volume={8},
  number={5},
  pages={054007},
  year={2017},
  publisher={APS}
}

@article{naaman2016chip,
  title={On-chip Josephson junction microwave switch},
  author={Naaman, Ofer and Abutaleb, MO and Kirby, Chris and Rennie, Michael},
  journal={Applied Physics Letters},
  volume={108},
  number={11},
  year={2016},
  publisher={AIP Publishing}
}

@article{wagner2019demonstration,
  title={Demonstration of a superconducting nanowire microwave switch},
  author={Wagner, Andrew and Ranzani, Leonardo and Ribeill, Guilhem and Ohki, TA},
  journal={Applied Physics Letters},
  volume={115},
  number={17},
  year={2019},
  publisher={AIP Publishing}
}

@article{sank2016measurement,
  title={Measurement-induced state transitions in a superconducting qubit: Beyond the rotating wave approximation},
  author={Sank, Daniel and Chen, Zijun and Khezri, Mostafa and Kelly, J and Barends, R and Campbell, B and Chen, Y and Chiaro, B and Dunsworth, A and Fowler, A and others},
  journal={Phys. Rev. Lett.},
  volume={117},
  number={19},
  pages={190503},
  year={2016},
  publisher={APS}
}

@article{macklin2015near,
  title={A near--quantum-limited Josephson traveling-wave parametric amplifier},
  author={Macklin, Chris and O’brien, K and Hover, D and Schwartz, ME and Bolkhovsky, V and Zhang, X and Oliver, WD and Siddiqi, I},
  journal={Science},
  volume={350},
  number={6258},
  pages={307--310},
  year={2015},
  publisher={American Association for the Advancement of Science}
}

@article{kow2026traveling,
  title={Traveling-wave parametric amplifier with passive reverse isolation},
  author={Kow, CS and Bell, MT},
  journal={Physical Review X},
  volume={16},
  number={2},
  pages={021003},
  year={2026},
  publisher={APS}
}

@article{malz2018quantum,
  title={Quantum-limited directional amplifiers with optomechanics},
  author={Malz, Daniel and T{\'o}th, L{\'a}szl{\'o} D and Bernier, Nathan R and Feofanov, Alexey K and Kippenberg, Tobias J and Nunnenkamp, Andreas},
  journal={Physical review letters},
  volume={120},
  number={2},
  pages={023601},
  year={2018},
  publisher={APS}
}

@article{lecocq2021efficient,
  title={Efficient qubit measurement with a nonreciprocal microwave amplifier},
  author={Lecocq, Florent and Ranzani, Leonardo and Peterson, Gabriel A and Cicak, Katarina and Jin, XY and Simmonds, RW and Teufel, JD and Aumentado, Jose},
  journal={Physical Review Letters},
  volume={126},
  number={2},
  pages={020502},
  year={2021},
  publisher={APS}
}

@article{stefanazzi2022qick,
  title={The QICK (Quantum Instrumentation Control Kit): Readout and control for qubits and detectors},
  author={Stefanazzi, Leandro and Treptow, Kenneth and Wilcer, Neal and Stoughton, Chris and Bradford, Collin and Uemura, Sho and Zorzetti, Silvia and Montella, Salvatore and Cancelo, Gustavo and Sussman, Sara and others},
  journal={Review of Scientific Instruments},
  volume={93},
  number={4},
  year={2022},
  publisher={AIP Publishing}
}

@article{ding2024experimental,
  title={Experimental advances with the QICK (Quantum Instrumentation Control Kit) for superconducting quantum hardware},
  author={Ding, Chunyang and Di Federico, Martin and Hatridge, Michael and Houck, Andrew and Leger, Sebastien and Martinez, Jeronimo and Miao, Connie and I, David Schuster and Stefanazzi, Leandro and Stoughton, Chris and others},
  journal={Physical Review Research},
  volume={6},
  number={1},
  pages={013305},
  year={2024},
  publisher={APS}
}

@article{anferov2024superconducting,
  title={Superconducting qubits above 20 GHz operating over 200 mK},
  author={Anferov, Alexander and Harvey, Shannon P and Wan, Fanghui and Simon, Jonathan and Schuster, David I},
  journal={PRX quantum},
  volume={5},
  number={3},
  pages={030347},
  year={2024},
  publisher={APS}
}

@article{zhao2026persistent,
  title={Persistent-current-biased and current-actuated switch for superconducting circuits},
  author={Zhao, Ziyi and Gurra, Eva and Vissers, Michael R and Lehnert, KW},
  journal={Physical Review Applied},
  volume={26},
  number={2},
  pages={024036},
  year={2026},
  publisher={APS}
}

\end{document}